\documentclass[aps,pre,twocolumn,showpacs,superscriptaddress,groupedaddress]{revtex4-1}
\pdfoutput=1
\usepackage{hyperref}
\usepackage{epsfig}
\usepackage{amsmath}
\usepackage{graphicx}
\usepackage{wrapfig}
\usepackage{dcolumn}
\usepackage{bm}
\usepackage{amssymb}
\usepackage{titlesec}
\usepackage{mathtools}
\usepackage{relsize}
\usepackage{cleveref}
\usepackage[usenames,dvipsnames]{color}
\usepackage{ulem}

\begin{document}
\title{Enhancement of large fluctuations to extinction in adaptive networks}
\author{Jason Hindes}
\affiliation{U.S. Naval Research Laboratory, Code 6792, Plasma Physics Division, Nonlinear Systems Dynamics Section, Washington, DC 20375}
\author{Leah B. Shaw}
\affiliation{Department of Mathematics, College of William and Mary, Williamsburg, VA 23187}
\author{Ira B. Schwartz}
\affiliation{U.S. Naval Research Laboratory, Code 6792, Plasma Physics Division, Nonlinear Systems Dynamics Section, Washington, DC 20375}

\begin{abstract}
During an epidemic individual nodes in a network may adapt their connections to reduce the chance of infection. A common form of adaption is avoidance rewiring, where a non-infected node breaks a connection to an infected neighbor, and forms a new connection to another non-infected node. Here we explore the effects of such adaptivity on stochastic fluctuations in the susceptible-infected-susceptible model, focusing on the largest fluctuations that result in extinction of infection. 
Using techniques from large-deviation theory, combined with a measurement of heterogeneity in the susceptible degree distribution at the endemic state, we are able to predict and analyze large fluctuations and extinction in adaptive networks. We find that in the limit of small rewiring there is a sharp exponential reduction in mean extinction times compared to the case of zero-adaption. Furthermore, we find an exponential enhancement in the probability of large fluctuations with increased rewiring rate, even when holding the average number of infected nodes constant. 

\end{abstract}
\maketitle

\section{\label{sec:INTRO} INTRODUCTION}
Network dynamics is an important topic of research spanning nearly every field of science and engineering\cite{Vespignani1:Book,Newman:Book,Barabasi:Book}. From the standpoint of statistical physics, the study of network structure and dynamics offers a way to broaden principles, since the field is formulated in general topological terms\cite{Dorogovtsev:RMP2008}. Moreover, network-based approaches are useful for applying statistical physics to social and technological problems where interactions and behaviors are often complex\cite{BarabasiNetworkTakeOver}. However, the usual mode of analysis in network science is to separate topology and dynamics. This approach has produced broad insights on the topological dependencies of many processes: from traditional critical phenomena\cite{Dorogovtsev:RMP2008} to synchronization\cite{Lou2014} and optimal control\cite{LiuRMP2016,Klickstein2017}. One of the distinctive features of complex systems, however, is adaption, where a feedback loop exists between network topology and dynamics -- producing an even wider spectrum of behaviors for adaptive complex networks\cite{Gross:Book}.     

An important class of adaptive dynamics occurs during the spread of infection through contact networks, where a network responds in some way to the presence of infection\cite{GrossPRL2006,PastorRMP}. For instance, adaption may come in the form of avoidance ``rewiring", where a non-infected node breaks a connection to an infected neighbor, and forms a new connection to another non-infected node in the network\cite{GrossPRL2006,Leah2008PRE}. Much is known about the dynamics of adaptive networks, including the existence of bistability between disease-free and endemic states generically\cite{GrossPRL2006,Juher2013,Bodo2017}, and network oscillations given special rewiring mechanisms and model parameters\cite{GrossEPL2008,Rogers2012,Simon2016}. 

An important theoretical and practical question is how to control infection in such networks, and possibly rid a network of infection altogether\cite{WangPhysRep2016}. Typically, model systems are analyzed in deterministic limits. The result is a stability map specifying regions where distinct dynamical behaviors are stable\cite{PastorRMP}. The question of how infection can be extinguished, then, becomes a matter of either tuning parameters to regions where disease-free states are stable, or adding controls that steer networks toward unstable states\cite{WangPhysRep2016}. However, in all real finite networks the dynamics are uncertain, both because contact processes (such as the spread of disease) are inherently stochastic\cite{AnderssonBook,KeelingBook,Rogers2012} and because real networks operate in uncertain environments\cite{KamenevPRL2008}. The resulting stochastic fluctuations eventually cause {\it extinction} of infection in finite networks with probability one in the infinite-time limit\cite{Assaf2010,AssafRev,Meerson2010,SchwartzJRS2011,Bovenkamp2015,Nasell:Book,Meerson2010}.Yet, theoretical methods do not exist for predicting extinction in adaptive contact networks. 

Recent progress has been made in predicting, analyzing, and controlling large dynamical fluctuations of the sort that lead to extinction or switching between distinct behaviors in static networks\cite{HindesExtin2016,Motter2015,Assaf2012,Hindes2017PRE,HindesSciRep2017,Lindley2014}. This work has leveraged the theory of large deviations and rare events that was originally developed to understand large fluctuations, LFs, in well-mixed and spatially homogeneous systems\cite{Dykman1994,AssafRev,SchwartzJST2009,MeersonPRER2011,Chaudhury2012}. One of the main results of large-deviation theory is that if a fluctuation from the usual dynamics is a rare event, then the process by which it occurs is captured by a path of maximum probability, and is describable by analytical mechanics\cite{Dykman1994,Friedlin:Book,SchwartzJRS2011,Assaf2010,Luchinsky1997}. This work continues the extension of large-deviation theory to adaptive networks, enabling us to capture the patterns of LFs causing extinction in such networks and quantify how their statistics depend on adaptive rewiring. The paper layout is the following: Sec.\ref{sec:Dynamics} introduces the model dynamics and examines the statistical properties of extinctions in adaptive networks as rare events, Sec.\ref{sec:WKB} outlines large-deviation techniques for predicting extinction in two parameter regions (corresponding to distinct types of extinction), and Sec.\ref{sec:Simulations} discusses the enhancement of LFs due to adaptive rewiring with several examples.  

\section{\label{sec:Dynamics} ADAPTIVE NETWORK DYNAMICS AND EXTINCTION}
In this work we consider an adaptive network with a fixed number of nodes, $N$, and edges, $N\!\left<k\right>\!/2$, where $\left<k\right>$ is the average degree of a node. The nodes represent individuals that are either infected or susceptible (non-infected). Susceptible nodes become infected at a rate proportional to the number of their infected neighbors in the network: $\beta$ per infected neighbor. Infected nodes recover spontaneously at a rate $\alpha$. In addition to the standard infection dynamics, edges between susceptible and infected nodes are rewired at a rate $w$ \cite{GrossPRL2006}. Rewiring entails a susceptible node breaking an edge with an infected neighbor and forming a new edge with another susceptible node chosen uniformly at random. Both multiple edges to neighbors and self-edges are excluded in rewiring. Since we are interested in the effects of adaptive rewiring on extinction (as opposed to other topological properties\cite{Hindes2017PRE}), we take $w\!=\!0$ to be an Erd\H{o}s-R\'{e}nyi network\cite{Newman:Book}.    

\begin{figure}[b]
\includegraphics[scale=0.238]{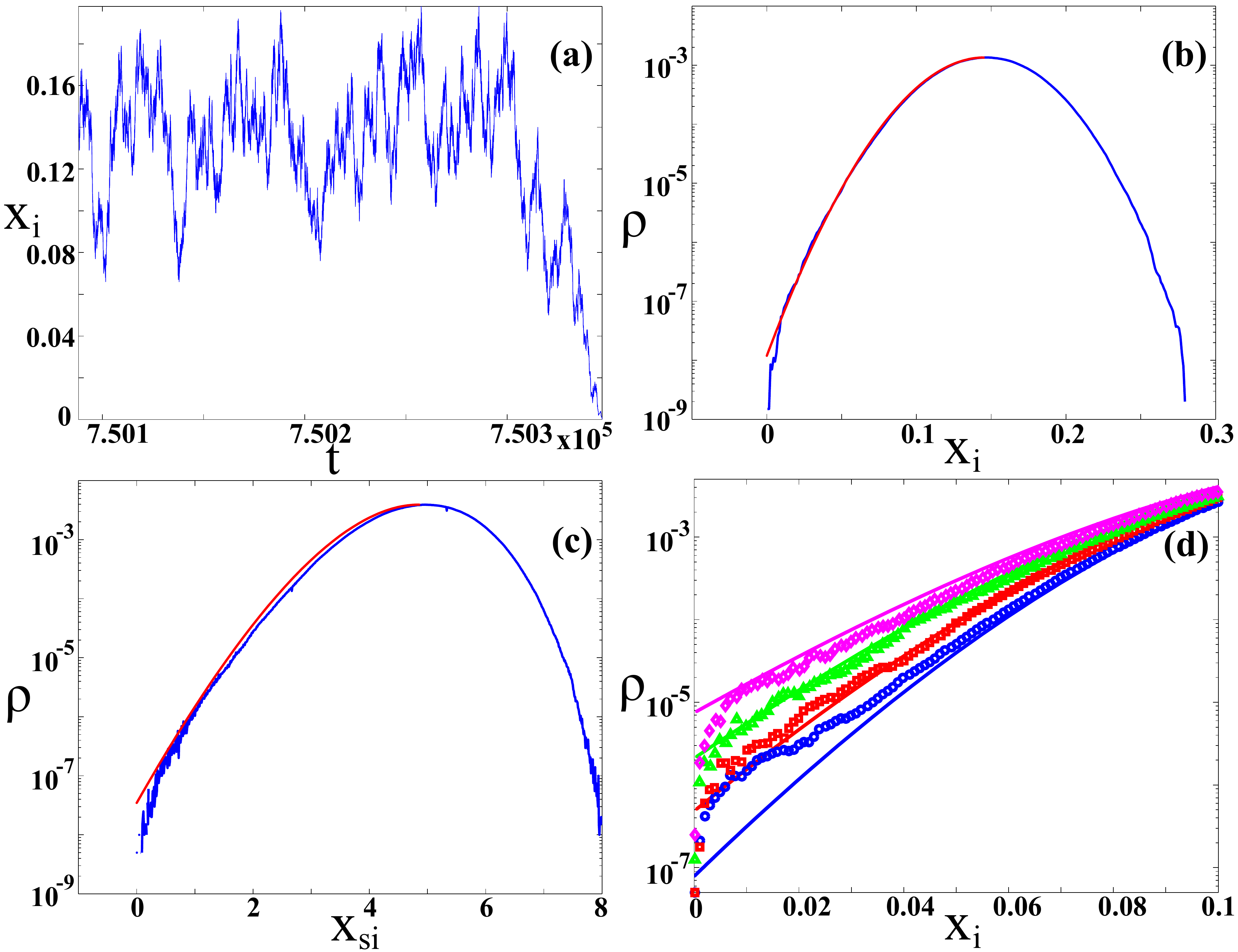}
\caption{(Color online) Large fluctuations in adaptive networks. (a) Fraction of infected nodes, $x_{i}$, at time $t$ in a single stochastic trajectory. (b) histogram of the trajectory, $\rho$, as a function of $x_{i}$, shown in blue. (c) histogram of per-capita number of susceptible-infected edges, $x_{si}$, shown in blue. Parameters for (a)-(c) are $N\!=\!1000$, $\alpha\!=\!1$, $\beta\!=\!0.03$, $w\!=\!0.03$, and $\left<k\right>\!=\!40$, for which the endemic state is the only deterministically stable state. Predictions from Eqs.(\ref{eq:WKB})-(\ref{eq:Action}) are shown in red for (b)-(c). (d) Histograms of $x_{i}$ for several $w$, and $x_{i}^{*}$ held constant. Thirty stochastic trajectories are averaged with parameters: ($w\!=\!0.10,\beta\!=\!0.03113$, blue $\circ$), ($w\!=\!0.25,\beta\!=\!0.033033$, red $\Box$), ($w\!=\!0.40,\beta\!=\!0.035051$, green $\triangle$), and  ($w\!=\!0.55,\beta\!=\!0.03711$, magenta $\diamondsuit$). Predictions are shown with solid lines; $N\!=\!1000$, $\alpha\!=\!1$, and $\left<k\right>\!=\!40$.}
\label{fig:Distribution}
\end{figure}

As defined by these stochastic reactions, the network dynamics is a continuous-time Markov process and is, therefore, inherently {\it noisy}. A single stochastic realization can be generated from a Gillespie algorithm\cite{Gillespie2013} starting from an initial number of infected nodes. In any finite adaptive network, the number of infected nodes and edges connected to infected nodes fluctuate in time around a mean non-zero number, or endemic state (assuming $\beta$ sufficiently large and initial conditions that relax to an endemic state). Over long time scales, LFs emerge in which the interplay between noise and network dynamics results in large deviations from the endemic state\cite{AssafRev}. In the most extreme cases, LFs cause a complete extinction of infection in the network, where all nodes become susceptible and the dynamics terminates\cite{SchwartzJST2009}. This process can be seen in Fig.\ref{fig:Distribution} (a), where the fraction of infected nodes in an adaptive network fluctuates in time from $t\!=\!0$ until an extinction occurs.

\begin{figure}[t]
\includegraphics[scale=0.232]{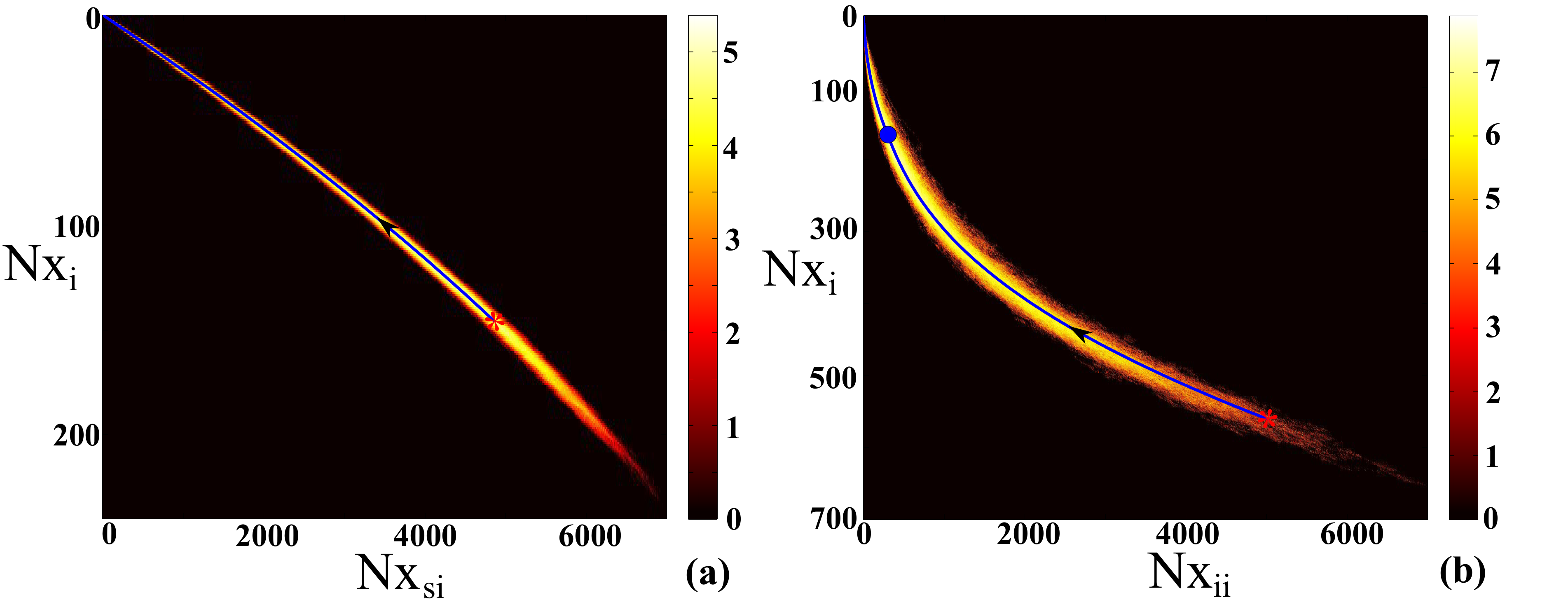}
\caption{(Color online) Two-dimensional projections of the density function of extinction paths computed from stochastic simulations, shown on log scale. (a) Number of infected nodes and the number of susceptible-infected edges are specified on the vertical and horizontal axes, respectively. Parameters are $N\!=\!1000$, $\alpha\!=\!1$, $\beta\!=\!0.03$, $w\!=\!0.03$, and $\left<k\right>\!=\!40$. (b) Number of infected nodes and the number of infected-infected edges are specified on the vertical and horizontal axes, respectively. Parameters are $N\!=\!1000$, $\alpha\!=\!1$, $\beta\!=\!0.103$, $w\!=\!5.10  $, and $\left<k\right>\!=\!40$. Predictions from Eq.(\ref{eq:Hamiltons}) are shown in blue with endemic states (red $*$) and a saddle (blue $\circ$).}
\label{fig:HeatMaps}
\end{figure}

When the network is large $(N\!\gg\!1)$ and/or the dynamics are sufficiently far from bifurcations, the probability distribution of network states is approximately time-independent, or {\it quasistationary}, with an exponential tail\cite{Nasell:Book,AssafRev}. The latter property is viewable in Fig.\ref{fig:Distribution} (b)-(c), which show histograms of time-series data from stochastic realizations of the adaptive network dynamics starting from an endemic state, e.g., Fig.\ref{fig:Distribution} (a). We can see that the very largest fluctuations from the endemic state (where the probability is a maximum) occur with exponentially small probabilities, and therefore, are called {\it rare events}. Moreover, when LFs are such rare events, they arise, effectively, through a unique dynamical sequence\cite{Dykman1994,SchwartzJRS2011}. We can demonstrate this fact by examining the density of many stochastic trajectories that end with extinction. The result is a narrow distribution that follows {\it a single path} of maximum probability, as shown in Fig.\ref{fig:HeatMaps}. It is our goal to analyze the statistics of such trajectories corresponding to LFs, and understand how LFs depend on the rate of rewiring in adaptive networks.  

\begin{figure}[t]
\includegraphics[scale=0.217]{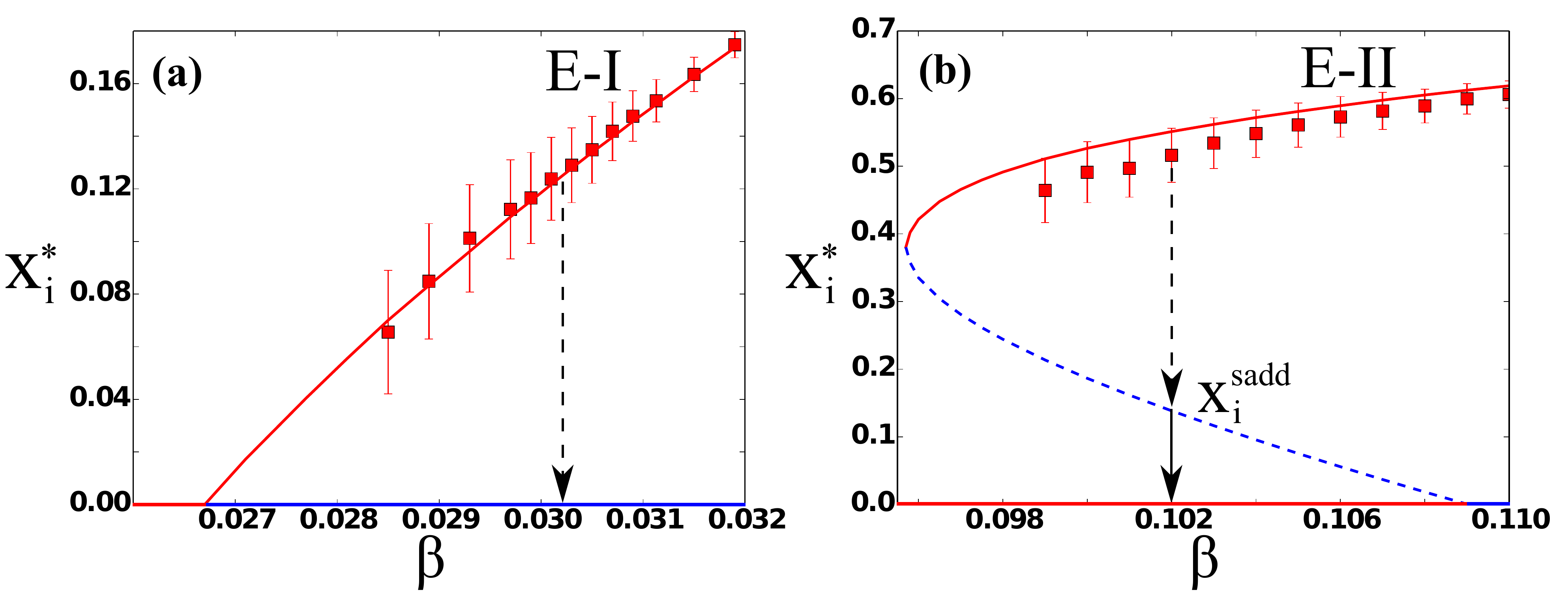}
\caption{(Color online) Extinction schematics shown with the fraction of infected nodes at the endemic state, $x_{i}^{*}$, versus infection rate. (a) E-I: large fluctuation (dashed-arrow) leads from a deterministically stable (red) endemic state, $x_{i}^{*}$, to a deterministically unstable (blue) state $x_{i}\!=\!0$. Red $\Box$'s are $\left<x_{i}\right>$ measured from simulations. Parameters are $N\!=\!1000$, $\alpha\!=\!1$, $w\!=\!0.10$, and $\left<k\right>\!=\!40$. (b) E-II: large fluctuation (dashed-arrow) leads from a deterministically stable endemic state to a saddle, $x_{i}^{sadd}$, (blue dashed line), after which the network follows the deterministic dynamics from $x_{i}^{sadd}$ to $x_{i}\!=\!0$ (solid arrow). Parameters are $N\!=\!1000$, $\alpha\!=\!1$, $w\!=\!5.0$, and $\left<k\right>\!=\!40$.}
\label{fig:Schematic}
\end{figure}

Though the model rules are simple, adaptive networks are known to produce a variety of dynamical behaviors that emerge through multiple bifurcations\cite{GrossPRL2006,Gross:Book,Juher2013,Bodo2017}. In this work we consider two regions of parameter space. In the first, which we call E-I (E stands for extinction), the endemic state is the only deterministically stable state and the disease-free state is unstable. Deterministic in this context refers to dynamics when $N\!\rightarrow\!\infty$. In the second, region E-II, both the disease-free and endemic states are deterministically stable and are separated by a saddle. The two cases imply qualitatively distinct types of LFs to extinction for networks with adaptivity, which are illustrated in Fig.\ref{fig:Schematic} \footnote{Error bars represent the standard deviation around the average in 200 simulations}.    

\section{\label{sec:WKB} PREDICTIONS WITH COMBINED PAIR AND WKB APPROXIMATIONS}
In order to predict LFs, we need to know the probability distribution for network states, $\rho$. In general, $\rho$ is governed by a high-dimensional master equation, describing the joint probability for all nodes in  all states\cite{KeelingBook,Dykman1994,PastorRMP}. This is far too much information to be practically useful for large adaptive networks. Therefore, we introduce a standard pair-approximation scheme\cite{Juher2013,GrossPRL2006,Rogers2012,Simon2016} to reduce the dimensionality, in which the network state, $\bold{X}$, describes only the total number of infected nodes $Nx_{i}$, susceptible-infected edges $Nx_{si}$, and infected-infected edges $Nx_{ii}$, or $\bold{X}\!=\!N(x_{i},x_{si},x_{ii})\!\equiv\!N\bold{x}$, where $\bold{x}$ is the per-capita value of $\bold{X}$. Since $N$ and $N\!\left<k\right>\!/2$ are fixed, $x_{i}\!+\!x_{s}\!=\!1$, and $x_{si}\!+\!x_{ii}\!+\!x_{ss}\!=\!\left<k\right>\!/2$, where $x_{ss}$ is the per-capita number of susceptible-susceptible edges. To find and solve an approximate master equation in terms of $\bold{X}$, we must know the rate for each reaction type, and how $\bold{X}$ changes when a reaction occurs\cite{Lindley2014,Rogers2012}. 

The three reactions are: rewiring (1), recovery (2), and infection (3) with rates 
\begin{align}
\label{eq:Rates}
R_{1}(\bold{X})&=wNx_{si},\nonumber \\
R_{2}(\bold{X})&=\alpha Nx_{i}, \nonumber \\ 
R_{3}(\bold{X})&=\beta N x_{si},
\end{align} 
respectively. The corresponding increments to $\bold{X}$ for each reaction are approximately, 
\begin{align}
\label{eq:Increments}
\boldsymbol{\delta}_{1}(\bold{x})&= (0,-1,0),\nonumber \\
\boldsymbol{\delta}_{2}(\bold{x})&=\Big(-1,\frac{2x_{ii}-x_{si}}{x_{i}},\frac{-2x_{ii}}{x_{i}}\Big), \nonumber \\ 
\boldsymbol{\delta}_{3}(\bold{x})&=\Big(1,\frac{z[\left<k\right>-3x_{si}-2x_{ii}]}{1-x_{i}}-1,\frac{zx_{si}}{1-x_{i}}+1\Big),
\end{align}
where $\bold{X}\!\rightarrow\!\bold{X}+\boldsymbol{\delta}_{j}$ for reaction-type $j$ (derivation of Eq.(\ref{eq:Increments}) in Sec.\ref{sec:Increments}). The parameter $z$ in $\boldsymbol{\delta}_{3}$ arises from the pair-approximation closure, and is a measure of heterogeneity in the susceptible degree distribution, $g_{k}^{s}$, where $g_{k}^{s}$ gives the fraction of susceptible nodes with degree $k$\cite{Juher2013}.  In particular, $z$ is the ratio of the average ``excess degree" of a susceptible node (the number of additional neighbors reachable from a node after following a randomly selected edge) to the average degree\cite{Newman:Book}. For networks without clustering and degree correlations    
\begin{align}
\label{eq:Z}
z\cong\sum_{k}g_{k}^{s}(k^{2}-k)\Big/\Big[\!\sum_{k}g_{k}^{s} k\Big]^{2}.  
\end{align}

A closed form expression is unknown for $z$ in adaptive networks. Typically, it is assumed that $z\!\equiv\!1$, which is equivalent to assuming that the susceptible degree distribution is a Poisson distribution\cite{GrossPRL2006,Juher2013,Gross:Book,Simon2016}. However, such an assumption is not sufficiently accurate for predicting LFs in adaptive networks. In this work, we {\it take $z$ to be constant in time along a large fluctuation} from the endemic state. The constant-$z$ assumption generalizes the Poisson assumption-- allowing for a more flexible degree distribution\cite{Juher2013}. In general, we find the approximation to be most accurate when rewiring is slow. 

In practice, the average value, $\left<z\right>$, is measured at the endemic state in simulations, and substituted into the equations below. By combining Eqs.(\ref{eq:Rates}-\ref{eq:Increments}) with the constant-$z$ assumption, we find an approximate master equation that can be analyzed to predict LFs:
\begin{align}
\label{eq:MasterEquation}
\frac{\partial \rho}{\partial t}(\bold{X},t)=\sum_{j=1}^{3}\!\Big[\rho(\bold{X}\!-\!\boldsymbol{\delta}_{j},t)R_{j}(\bold{X}\!-\!\boldsymbol{\delta}_{j})-\rho(\bold{X},t)R_{j}(\bold{X})\!\Big].   
\end{align}

As hinted by Fig.\ref{fig:Distribution}, the tail of the adaptive network's probability distribution, $\rho(\bold{X},t)$, is relevant for LFs\cite{Hindes2017PRE}. Therefore, we seek a special exponential solution of Eq.(\ref{eq:MasterEquation}) in the Wentzel-Kramers-Brillouin (WKB) form
\begin{align}
\label{eq:WKB}
\rho(\bold{X},t)=a\exp{\!\big\{\!-\!NS(\bold{x},t)\big\}}, 
\end{align}
(a standard ansatz from large-deviation theory)\cite{AssafRev,Assaf2010}. Substituting Eq.(\ref{eq:WKB}) into Eq.(\ref{eq:MasterEquation}) and expanding in powers of $1/N$ gives a Hamilton-Jacobi equation\cite{Dykman1994} at lowest order for the probability exponent, $S(\bold{x},t)$: 
\begin{align}
\label{eq:HJE}
\frac{\partial S}{\partial{t}} + H\Big(\bold{x},\frac{\partial S}{\partial\bold{x}}\;\Big) =0.
\end{align}
The functions $S(\bold{x},t)$ and $H(\bold{x},\partial S/\partial\bold{x})$ are called the action and Hamiltonian respectively (familiar from analytical mechanics\cite{AssafRev}). Solving Eq.(\ref{eq:HJE}) for the action gives the probability distribution exponent for LFs, Eq.(\ref{eq:WKB})\cite{Meerson2010}. In fact, since the distribution of LFs is quasistationary, Eq.(\ref{eq:HJE}) has the special zero-energy form, $\partial S/\partial t\!=\!H\!=\!0$ for all $t$.  

In order to find solutions of Eq.(\ref{eq:HJE}), it is useful to continue the analogy with analytical mechanics, and introduce a conjugate momentum, $\bold{p}\!\equiv\!\partial S/\partial\bold{x}$\cite{KamenevPRL2008}. In terms of $\bold{p}$, the Hamiltonian for the adaptive network is  
\begin{align}
\label{eq:Hamiltonian}
&H(\bold{x},\bold{p})=\sum_{j=1}^{3}\!\big[R_{j}(\bold{X})/N\big]\!\big[\!\exp{\!\big\{\bold{p}\cdot\boldsymbol{\delta}_{j}(\bold{x})\big\}}-1\big].  
\end{align}
From the Hamiltonian the network action can be found by solving Hamilton's equations of motion:
\begin{align}
\label{eq:Hamiltons}
\dot{\bold{x}}=\frac{\partial H}{\partial\bold{p}}, \;\text{and}\; \dot{\bold{p}}=-\frac{\partial H}{\partial\bold{x}},  
\end{align}
which are given explicitly in Sec.\ref{sec:Computation}. {\it The solutions of Eq.(\ref{eq:Hamiltons}) that correspond to LFs and extinction are unique forward-in-time $(\bold{x},\bold{p})$ trajectories connecting endemic and disease-free states with $H\!=\!0$}. Once such trajectories are computed, the action is simply the line integral of the momentum:  
\begin{align}
S(\bold{x})=\int_{\bold{x}^{*}}^{\bold{x}}\!\!\bold{p}(\bold{x}')\cdot{d\bold{x}'}, 
\label{eq:Action}
\end{align}
where $\bold{x}^{*}$ is the endemic state. 

It is important to note that solutions of Eqs.(\ref{eq:HJE}-\ref{eq:Action}) {\it minimize} the action, and therefore correspond to most probable, or optimal, paths of LFs\cite{Dykman1994}. Optimal paths, OPs, of Eq.(\ref{eq:Hamiltons}) can be determined numerically from boundary conditions with standard quasi-Newton methods\cite{Lindley2013}. In general, time-derivatives vanish at all boundaries, $(\dot{\bold{x}},\dot{\bold{p}})\!=\!(\bold{0},\bold{0})$, though the precise boundary conditions for Eq.(\ref{eq:Hamiltons}) depend on whether the disease-free state is deterministically stable or not\cite{AssafRev,Assaf2010}. For E-I the OP is a heteroclinic orbit of Eq.(\ref{eq:Hamiltons}) from an endemic state $(\bold{x}\!=\!\bold{x}^{*},\bold{p}\!=\!\bold{0})$ to a disease-free state {\it with negative momentum} $(\bold{0},\bold{p}^{*})$. The negative-momentum boundary condition is viewable in Fig.\ref{fig:Distribution} (b)-(d) where $\ln{\rho}$ has a positive slope at $\bold{x}\!=\!\bold{0}$. Extinction in region E-I is the generic form for static networks\cite{Hindes2017PRE}.  

In contrast for E-II, both the endemic and disease-free states are deterministically stable. When rewiring is fast ($w\!\gtrsim\!\alpha$) and infection levels are small, infected nodes become isolated and recover before infecting susceptible nodes. The result is extinction, even without LFs. However, when infection levels are high, frequent rewiring leads to ``clusters'' of high degree susceptibles\cite{GrossPRL2006} that, once infected, have a greater potential to spread the disease further. In this case, infection is sustainable. Between the two cases, there is a saddle point, at which slightly more infected nodes will produce a growth of infection, and slightly fewer infected nodes will result in extinction\cite{GrossPRL2006,Juher2013,Gross:Book}. Intuitively then, the most likely path to extinction from the endemic state is a LF directly to the saddle, since once the saddle is reached the network can follow the deterministic dynamics to extinction. Our approach formalizes this intuition. We find that the OP is a heteroclinic connection from $(\bold{x}^{*},\bold{0})$ to $(\bold{x}^{sadd},\bold{0})$. In E-II, extinction is analogous to noise-induced network switching between deterministically stable states\cite{HindesSciRep2017,Dykman1994}. More details about computing OPs are given in Sec.\ref{sec:Computation} and in \footnote{Example \uppercase{MATLAB} code is given in the supplementary material of the published version}.  


\section{\label{sec:Simulations} ENHANCEMENT OF LARGE FLUCTUATIONS}
Comparisons between simulations and predictions in Figs.\ref{fig:Distribution}-\ref{fig:Schematic} were made by measuring $\left<z\right>$ at equilibrium, using Eq.(\ref{eq:Z}), and computing the network OPs and actions from Eq.(\ref{eq:Hamiltons}-\ref{eq:Action}) with $z\!=\!\left<z\right>$. As expected, both probability and OP predictions are in good quantitative agreement for moderate rewiring rates $(w\!<\!\alpha)$ (see Fig.\ref{fig:Distribution} and Fig.\ref{fig:HeatMaps} (a)). However, even for fast rewiring $(w\!\gtrsim\!\alpha)$, i.e., E-II, OP predictions capture the qualitative shape of LFs, as demonstrated in Fig.\ref{fig:HeatMaps} (b).     

\begin{figure}[h]
\includegraphics[scale=0.231]{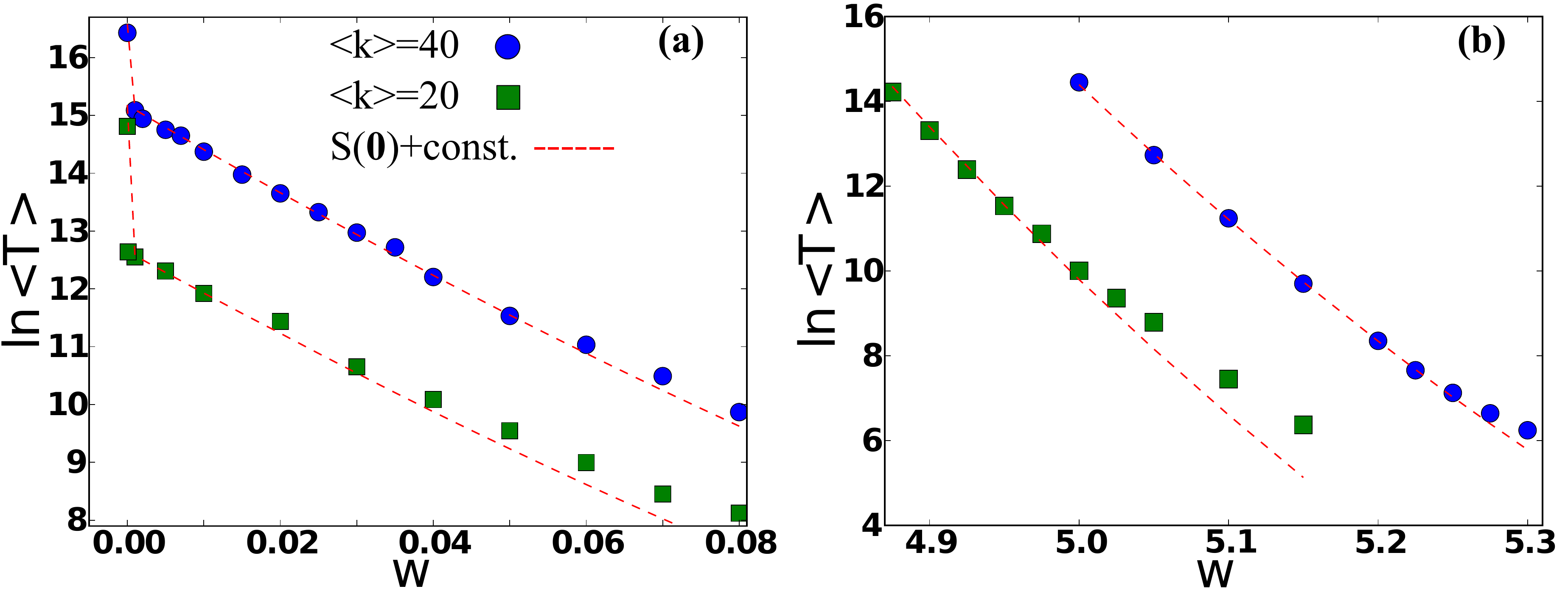}
\caption{(Color online)  Log of the mean extinction time vs. rewiring rate. Mean times were taken from $400$ stochastic realizations of the dynamics. A single Erd\H{o}s-R\'{e}nyi network was used as an initial condition for each series with a uniformly-random distribution of infection such that $x_{i}\!=\!x_{i}^{*}$. $N\!=\!1000$ and $\alpha\!=\!1$ for all networks, with average degrees: $\left<k\right>\!=\!40\;(\circ)$ and $\left<k\right>\!=\!20\;(\Box)$. (a) E-I: $\beta\!=\!0.03\;(\circ)$ and $\beta\!=\!0.06\;(\Box)$. (b) E-II: $\beta\!=\!0.103\;(\circ)$ and $\beta\!=\!0.206\;(\Box)$. Predictions from Eqs.(\ref{eq:Action}-\ref{eq:Time}) are shown in red (dashed). The constant in Eq.(\ref{eq:Time}) was fit to match simulations.}
\label{fig:Times}
\end{figure}

In addition to the OP and probabilities of LFs, we can use the WKB method to approximate the mean lifetime of the endemic state, which measures how long it takes on average for an adaptive network to reach extinction when starting from the endemic state. If we assume that the rate, or inverse mean time $1/\!\left<T\right>$, of extinction is proportional to the probability of extinction\cite{Assaf2010}, we expect  
\begin{align}
\label{eq:Time}
\ln{\left<T\right>}\cong S(\bold{0})+\text{const.} 
\end{align}
Predictions from Eq.(\ref{eq:Time}) are shown in Fig.\ref{fig:Times} for E-I and E-II, and are in good quantitative agreement with simulations. As a general feature of both extinction scenarios, we find large exponential decreases in the average extinction time, with small increases in $w$. The accuracy of extinction-time predictions for fast rewiring is somewhat surprising, since predictions for $x_{i}^{*}$ are noticeably off (see Fig.\ref{fig:Schematic} (b)). As a consequence, predictions for $\rho(\bold{x})$ are inaccurate near the endemic state for fast rewiring.

Since rewiring necessarily inhibits the local spread of infection, we may wonder whether the increased probability of LFs to extinction with increasing $w$ is merely due to a reduction in infection or whether it reflects an increase in stochastic effects. To isolate LFs from average infection, we compare $\rho(\bold{x})$ for different $w$ and $\beta$ chosen such that $Nx_{i}^{*}$ is a constant. Four examples are compared in Fig.\ref{fig:Distribution} (d) in E-I. It is demonstrated that even though $\rho(\bold{x})$ has a maximum at the same total number of infected nodes for each network, the relative {\it probabilities for fluctuations are exponentially larger in networks with higher rewiring rates}, e.g., a fivefold increase in $w$ results in a hundredfold increase in the probability of extinction, $\rho(\bold{0})$, for $N\!=\!1000$ and $x_{i}^{*}\!=\!0.15$.     

\begin{figure}[h]
\includegraphics[scale=0.243]{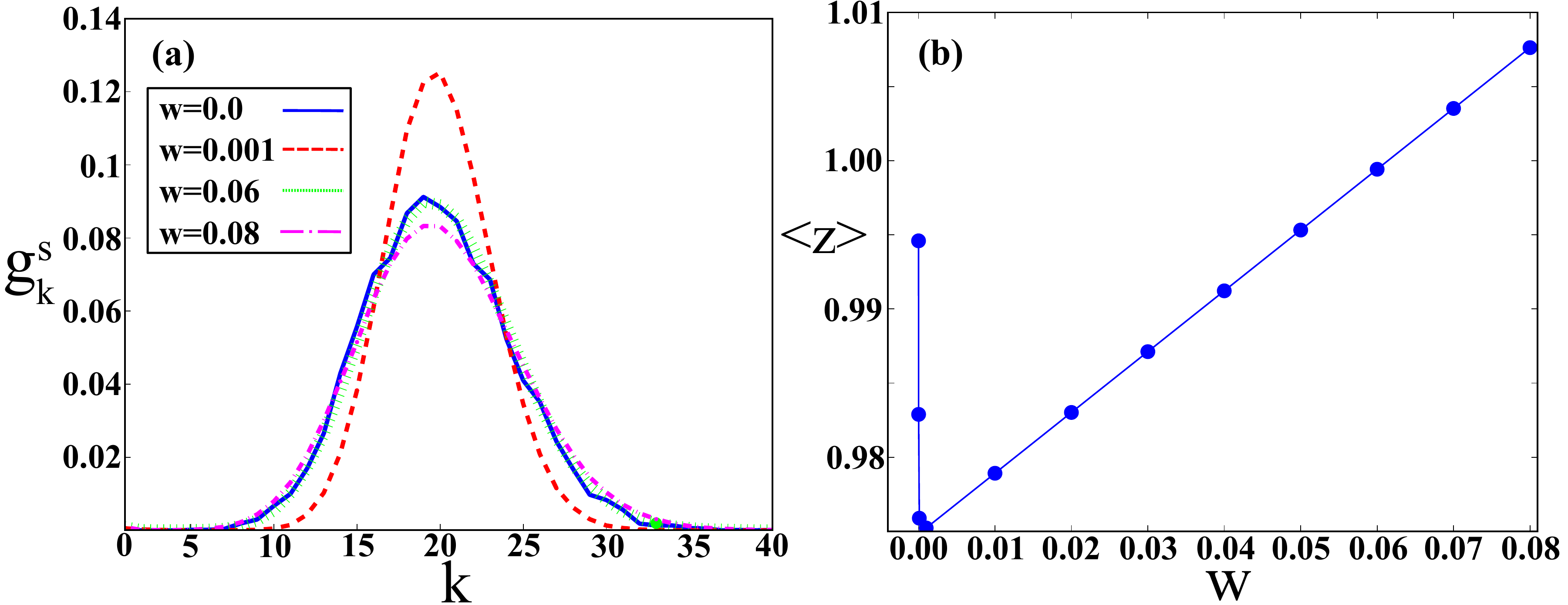}
\caption{(Color online) Average susceptible degree distribution for several rewiring rates in E-I. The distribution is an average of $200$ stochastic realization at $t\!\sim\!\left<T\right>\!$, with initial conditions chosen as in Fig.\ref{fig:Times}. (a) Fraction of susceptible nodes with degree $k$, $g_{k}^{s}$, vs. $k$. (b) $\left<z\right>$ vs. $w$. Parameters are $N\!=\!1000$, $\alpha\!=\!1$, $\left<k\right>\!=\!20$, and $\beta\!=\!0.06$.}
\label{fig:DD}
\end{figure}
In addition, a {\it sharp reduction} in extinction times with very small rewiring rates is accurately predicted for E-I, as demonstrated in Fig.\ref{fig:Times} (a). The reduction is a consequence of the susceptible degree distribution becoming more homogeneous for $0\!\lesssim\!w\!\ll\!\beta$, and is not predicted by the usual $z\!\equiv\!1$ assumption. To explain the reduction in extinction times, we first consider the susceptible degree distributions shown in Fig.\ref{fig:DD} (a) for several $w$. Figure \ref{fig:DD} (b) gives the values of $\left<z\right>$. We find that the susceptible degree distribution narrows as $w$ is increased from zero for very small $w$, and then broadens as $w\!\rightarrow\!\beta$. The narrowing of the degree distribution when $0\!\lesssim\!w\!\ll\!\beta$ occurs because high degree nodes are more likely to be infected, and thus more likely to have their susceptible neighbors rewire away. In contrast, lower degree nodes are more likely to be susceptible and thus to accumulate more neighbors when randomly chosen as targets for rewiring. Therefore, {\it for small rewiring the number of nodes with extreme degree is reduced}, with more nodes brought nearer the mean (Note: the large peak in the susceptible degree distribution around the average degree for $w\!=\!0.001$ in Fig.\ref{fig:DD} (a)).  
The result is a reduction in $z$. For comparison, $z\!=\!1-1/\!\left<k\right>$ for a perfectly homogenous degree distribution.  

The sharp exponential reduction in extinction times with small rewiring can be understood in the following way: it is known that heterogeneous networks have lower epidemic thresholds for a given $\beta$ \cite{PastorRMP}. For instance in the model considered here, the critical infection rate at which the disease-free state changes stability is $\beta_{cr}\!=\!(\alpha+w)/[z\!\left<k\right>]$\cite{Juher2013}. Hence, if $z$ is reduced at constant $\beta>\beta_{cr}$ for $0\!\lesssim\!w\!\ll\!\beta$, as occurs in Fig.\ref{fig:DD} (b), $\beta_{cr}$  increases and the network gets closer to the epidemic threshold. The decrease in the distance from threshold results in an the exponential increase in $\rho(\bold{0})$ and decrease in $\left<T\right>$\cite{Assaf2010,HindesExtin2016,Hindes2017PRE}, as in Fig.\ref{fig:Times} (a) for $w\!\gtrsim\!0$.    

When rewiring is fast, however, infected nodes are disproportionately low degree, since they shed their edges quickly. When such infected nodes recover, they become susceptible nodes with low degree. Conversely, nodes that have been susceptible for a long time, have accumulated many neighbors and therefore higher degree. Combining the groups of susceptible nodes produces a broadening of the susceptible degree distribution (after the initial narrowing for small $w$) and an increase in $z$ with increasing $w\!\sim\!\beta$ \cite{GrossPRL2006}, shown in Fig.\ref{fig:DD} (b).
\begin{figure}[t]
\includegraphics[scale=0.27]{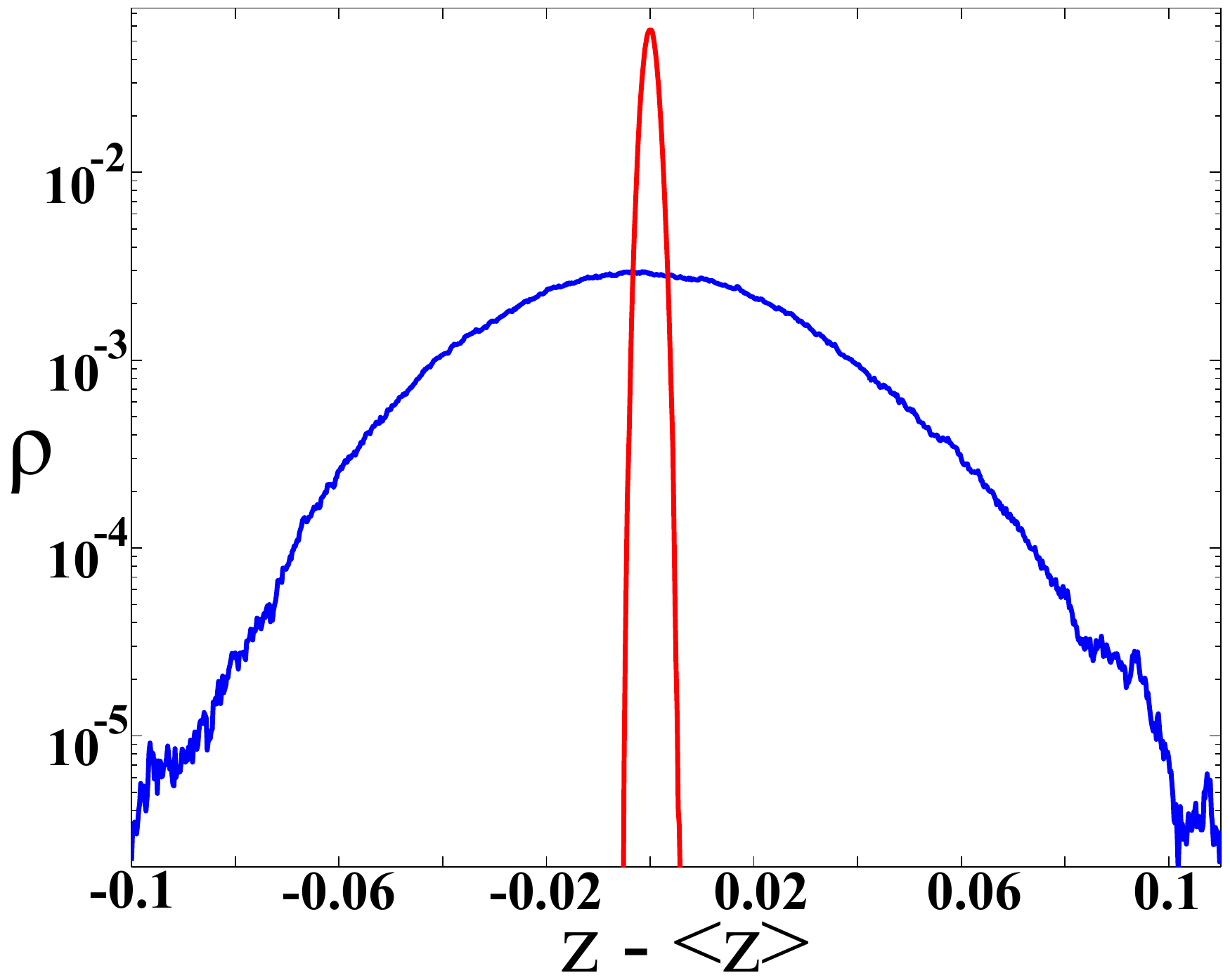}
\caption{(Color online) Histogram of $z-\left<z\right>$ for slow (red) and fast (blue) rewiring, Eq.(\ref{eq:Z}). The parameters are $N\!=\!1000$, $\left<k\right>\!=\!40$, and $\alpha=1$. Slow rewiring: $\beta\!=\!0.03$ and $w\!=\!0.06$. Fast rewiring: $\beta\!=\!0.103$ and $w\!=\!5.0$.}
\label{fig:Zchange}
\end{figure} 

Yet when $w\!\gg\!\alpha$ the constant-$z$ and negligible degree-correlations\cite{GrossPRL2006} assumptions are no longer accurate, as LFs in the dynamics induce LFs in the network topology beyond the mean connectivity between types of nodes that is treated explicitly by our approach. The variation in $z$ with fast rewiring is depicted in Fig.\ref{fig:Zchange}, which shows histograms of $z-\left<z\right>$ for slow (red) and fast (blue) rewiring examples. In the former, $z$ is sharply peaked around its mean value with a standard deviation of $0.001$. In the latter, the variation is much larger: e.g., a standard deviation of $0.03$ for the parameters given in Fig.\ref{fig:Zchange}. Moreover, the fluctuations in $z$ grow as bifurcations are approached. The variation in $z$ produces quantitative errors in predictions (e.g., the aforementioned mismatch between $x_{i}^{*}$ and $\left<x_{i}\right>$ shown in Fig.\ref{fig:Schematic}(b)). 

\section{\label{sec:Conclusion} CONCLUSION}
This work dealt with stochastic fluctuations in the susceptible-infected-susceptible model on finite networks with adaptive rewiring. By combining large-deviation and pair-approximation techniques with a measurement of heterogeneity in the susceptible degree distribution, we were able to predict and analyze large fluctuations that cause extinction in adaptive networks. Our approach allowed us to accurately predict the optimal path and statistics of extinction from deterministically stable endemic states for two distinct scenarios: extinction to an unstable disease-free state and extinction to a stable disease-free state through a saddle. 

In general, we have shown that adaptivity significantly alters the probabilities and optimal path of finite-size fluctuations in networks. An important finding was the exponential increase in the probabilities of large fluctuations with increased rewiring rate, even when the number of infected nodes at equilibrium is held constant. As a practical consequence, if epidemiological data are available only on the average number of individuals infected in a population, our work demonstrates that it is not sufficient to predict the probability of extinction or mean extinction times, since predictions are exponentially sensitive to the amount of adaptive avoidance in a population. Moreover, we have shown that even a very small amount of rewiring in networks can result in exponential reductions in extinction times when compared to networks without rewiring. The latter occurs because adaptivity causes networks to become more homogeneous in their degree distributions for slow rewiring rates. Future work will explore the use of adaptivity as a control strategy when combined with the stochastic effects considered.

In addition, we found that large fluctuations occur in topological structure not treated explicitly by pair approximation, resulting in quantitative errors. Analysis of this structure is yet lacking, even for deterministic limits\cite{Juher2013}. As this work demonstrates, however, deterministic approaches can be naturally extended to predict large fluctuations when combined with WKB techniques. Therefore, once a more accurate technique is found for analyzing adaptive network dynamics, the problem of fully predicting and analyzing large fluctuations can be solved. 

\section{\label{sec:Ack}ACKNOWLEDGMENTS}
\noindent JH is a National Research Council postdoctoral fellow. IBS was supported by the U.S. Naval Research Laboratory funding (Grant No. N0001414WX00023) and office of Naval Research (Grants No N0001416WX00657 and No. N0001416WX01643). LBS was supported by NSF Grant No DMS-1715651. This work was performed in part using computing facilities at the College of William and Mary which were provided by contributions from the National Science Foundation, the Commonwealth of Virginia Equipment Trust Fund and the Office of Naval Research. 

\appendix
\section{\label{sec:Appendix} APPENDIX}
\subsection{\label{sec:Increments} Increments} 
As mentioned in Sec.\ref{sec:WKB}, in order to find a master equation for the adaptive network, we must know how $\bold{X}\!=\!N(x_{i},x_{si},x_{ii})$ changes when reactions occur, Eq.(\ref{eq:Increments}). We call these reaction increments. Our strategy is to approximate the increments by their expected values in the network. The first reaction is rewiring, in which the number of infected nodes and infected-infected edges is constant, and the number of susceptible-infected edges is reduced by $1$. Therefore, $\bold{X}\!\rightarrow\!\bold{X}+\!(0,-1,0)$ when rewiring occurs, giving $\boldsymbol{\delta}_{1}$ in Eq.(\ref{eq:Increments}). 

The second reaction is recovery, where the number of infected nodes is reduced by $1$. Moreover, all of the old infected-infected edges attached to the recovered node become susceptible-infected edges and the old susceptible-infected edges become susceptible-susceptible edges. The expected numbers of old infected-infected edges and susceptible-infected edges connected to the newly susceptible node are $2x_{ii}/x_{i}$ and $x_{si}/x_{i}$, respectively. Hence when a recovery occurs 
\begin{equation}
\bold{X}\!\rightarrow\!\bold{X}\!+\!(-1,[2x_{ii}\!-\!x_{si}]/x_{i},-2x_{ii}/x_{i}), \nonumber
\end{equation}
giving $\boldsymbol{\delta}_{2}$ in Eq.(\ref{eq:Increments}).

The last reaction is infection, where the number of infected nodes increases by $1$, the old susceptible-infected edges attached to the newly infected node become infected-infected edges, and the old susceptible-susceptible edges become susceptible-infected edges. In order to find the expected increment for infection, we need to know the expected number of neighbors of a susceptible node that are reachable after following a randomly selected edge (namely the infected-susceptible edge along which the susceptible node was infected). As is customary, we refer to this number as $q_{s}$\cite{Juher2013,Newman:Book}. The excess degree of a susceptible node is therefore $q_{s}\!-\!1$. Given the average degree for a susceptible node, $\left<k\right>_{\!s}$, the parameter $z$ is defined such that $z\!\equiv\![q_{s}\!-\!1]/\!\left<k\right>_{\!s}$. Hence, the expected number of old susceptible-infected edges connected to the newly infected node is: the excess degree, $q_{s}\!-\!1$, multiplied by the fraction of edges that connect to infected neighbors from a susceptible node, $x_{si}/[(1\!-\!x_{i})\!\left<k\right>_{\!s}]$, plus one, or $(q_{s}\!-\!1)x_{si}/[(1\!-\!x_{i})\!\left<k\right>_{\!s}]\!+\!1$. Similarly, the expected number of old susceptible-susceptible edges is $(q_{s}\!-\!1)\!*\!2x_{ss}/[(1\!-x_{i})\!\left<k\right>_{\!s}]$. Putting these together we get 
\begin{equation}
\bold{X}\rightarrow\bold{X}\!+\!(1,z[2x_{ss}\!-\!x_{si}]/[1\!-\!x_{i}]\!-\!1,zx_{si}/[1\!-\!x_{i}]+1), \nonumber
\end{equation}
giving $\boldsymbol{\delta}_{3}$ in Eq.(\ref{eq:Increments})

\subsection{\label{sec:Computation} Optimal path computation}
Hamilton's equations of motion derive from the Hamiltonian, Eq.(\ref{eq:Hamiltonian}). Substituting Eq.(\ref{eq:Hamiltonian}) into Eq.(\ref{eq:Hamiltons}) gives: 
\begin{align}
\label{eq:EOM1}
\dot{x}_{i}=&\beta x_{si}e^{\bold{p}\cdot\boldsymbol{\delta_{3}}} -\alpha x_{i}e^{\bold{p}\cdot\boldsymbol{\delta_{2}}},\\
\label{eq:EOM2}
\dot{x}_{si}=&\beta x_{si}\Big[z\frac{\left<k\right>-3x_{si}-2x_{ii}}{1-x_{i}}-1\Big]e^{\bold{p}\cdot\boldsymbol{\delta_{3}}}+\\
&\alpha(2x_{ii}-x_{si})e^{\bold{p}\cdot\boldsymbol{\delta_{2}}}-wx_{si}e^{\bold{p}\cdot\boldsymbol{\delta_{1}}},\nonumber\\
\label{eq:EOM3}
\dot{x}_{ii}=&\beta x_{si}\Big[1+\frac{zx_{si}}{1-x_{i}}\Big]e^{\bold{p}\cdot\boldsymbol{\delta_{3}}}-2\alpha x_{ii}e^{\bold{p}\cdot\boldsymbol{\delta_{2}}}\\
\label{eq:EOM4}
-\dot{p}_{i}=&\alpha\Big[\!\!-\!1+e^{\bold{p}\cdot\boldsymbol{\delta_{2}}}\Big\{\!1+\frac{p_{si}(x_{si}\!-\!2x_{ii})\!+\!2p_{ii}x_{ii}}{x_{i}}\!\Big\}\!\Big]\\
&\!+\!\frac{\beta zx_{si}}{(1-x_{i})^{2}}e^{\bold{p}\cdot\boldsymbol{\delta_{3}}}\!\Big[p_{si}\big(\!\!\left<k\right>-3x_{si}\!-\!2x_{ii}\big)\!+\!x_{si}p_{ii}\Big] \nonumber\\
\label{eq:EOM5}
-\dot{p}_{si}=&w\Big[e^{\bold{p}\cdot\boldsymbol{\delta_{1}}}\!-\!1\Big]\!-\!\alpha p_{si}e^{\bold{p}\cdot\boldsymbol{\delta_{2}}}+\\
&\beta\Big[\!-1\!+\!e^{\bold{p}\cdot\boldsymbol{\delta_{3}}}\Big\{\!1+\frac{zx_{si}\big(p_{ii}\!-\!3p_{si}\big)}{1-x_{i}}\!\Big\}\!\Big]\nonumber \\
\label{eq:EOM6}
-\dot{p}_{ii}=&2\alpha\big[p_{si}-p_{ii}\big]e^{\bold{p}\cdot\boldsymbol{\delta_{2}}}\!-\!\frac{2\beta z x_{si}p_{si}}{(1-x_{i})}e^{\bold{p}\cdot\boldsymbol{\delta_{3}}}. 
\end{align}

Examining Eqs.(\ref{eq:EOM1}-\ref{eq:EOM6}), we note that there is a potential indeterminate form $(0/0)$ if $\bold{x}\!\rightarrow\!\bold{0}$. In practice, $\bold{x}\!=\!\bold{0}$ only arises when computing OPs in E-I, since in E-II the path terminates at a saddle point, $\bold{x}^{sadd}\!\neq\!\bold{0}$. Nevertheless, the ratios $x_{si}/x_{i}$ and $x_{ii}/x_{i}$ in Eqs.(\ref{eq:EOM1}-\ref{eq:EOM6}) are finite along the OP, and we can avoid numerical sensitivity by defining $f_{s}\!\equiv\!x_{si}/x_{i}$ and $f_{i}\!\equiv\!x_{ii}/x_{i}$. In E-I, we compute the OP by replacing $x_{si}$ and $x_{ii}$ with $f_{s}x_{i}$ and $f_{i}x_{i}$. The auxiliary dynamics are $\dot{f_{s}}\!=[\dot{x}_{si}-f_{s}\dot{x}_{i}]/x_{i}$ and $\dot{f_{i}}\!=[\dot{x}_{ii}-f_{i}\dot{x}_{i}]/x_{i}$, with boundary conditions $\dot{f_{s}}\!=\!\dot{f_{i}}\!=\!0$ at both the endemic and extinct states. After solutions are computed in $f_{s}$ and $f_{i}$, given the boundary conditions, $x_{si}$ and $x_{ii}$ can be found from the above definitions. 

In order to determine an OP numerically with quasi-Netwon methods\cite{Lindley2013}, an initial trial solution is needed that is sufficiently accurate to allow for convergence. Our approach is to use the approximate form of the OP near bifurcation. Two important cases are: E-I near a transcritical bifurcation, and E-II near a saddle-node bifurcation\cite{Bodo2017}. We parameterize the trial solution in both cases in terms of a unit-length, $h$, as in\cite{Hindes2017PRE,HindesSciRep2017}. For the transcritical case the OP has a linear form:  
\begin{align}
\label{eq:Trans}
&x_{i}\approx x_{i}^{*}h, \;\;\;\;\;\;\;\;\;\;\;\;\;\;\;\;\;\;\;\;\; p_{i}\approx p_{i}^{e}(1\!-\!h) \nonumber \\
&f_{s}\approx f_{s}^{*}h+f_{s}^{e}(1\!-\!h), \;\; p_{si}\approx p_{si}^{e}(1\!-\!h) \nonumber \\
&f_{i}\approx f_{i}^{*}h+f_{i}^{e}(1\!-\!h), \;\;\; p_{ii}\approx p_{ii}^{e}(1\!-\!h).
\end{align}
As defined, $h\!=\!1$ is the endemic state (superscript $*$), satisfying $p_{i}\!=\!p_{si}\!=\!p_{ii}\!=\!\dot{x}_{i}\!=\!\dot{f}_{s}\!=\!\dot{f}_{i}\!=\!0$, and similarly, the extinct state (superscript $e$) satisfies $\dot{p}_{i}\!=\!\dot{p}_{si}\!=\!\dot{p}_{ii}\!=\!\dot{f}_{s}\!=\!\dot{f}_{i}\!=\!x_{i}\!=\!0$.

On the other hand, near a saddle-node bifurcation the momenta are quadratic functions of $h$. In general the OP depends on the eigenvector of the Jacobian for the linearized Eqs.(\ref{eq:EOM1}-\ref{eq:EOM6}) associated with the bifurcation. If we define the components at the endemic state as $\eta_{x_{i}}^{*}$, $\eta_{p_{i}}^{*}$, $\eta_{p_{si}}^{*}$, and $\eta_{p_{ii}}^{*}$ associated with $x_{i}$, $p_{i}$, $p_{si}$, and $p_{ii}$, respectively, the OP near the saddle-node is: 
\begin{align}
\label{eq:Sadd}
x_{i}&\approx x_{i}^{*}h+x_{i}^{sadd}(1\!-\!h), \;\;\; p_{i}\approx\!\frac{\eta_{p_{i}}^{*}}{\eta_{x_{i}}^{*}}[x_{i}^{sadd}-x_{i}^{*}]h(1\!-\!h) \nonumber \\
x_{si}&\approx x_{si}^{*}h+x_{si}^{sadd}(1\!-\!h), \; p_{si}\approx\!\frac{\eta_{p_{si}}^{*}}{\eta_{x_{i}}^{*}}[x_{i}^{sadd}-x_{i}^{*}]h(1\!-\!h) \nonumber \\
x_{i}&\approx x_{ii}^{*}h+x_{ii}^{sadd}(1\!-\!h), \;\; p_{ii}\approx\!\frac{\eta_{p_{ii}}^{*}}{\eta_{x_{i}}^{*}}[x_{i}^{sadd}-x_{i}^{*}]h(1\!-\!h).
\end{align}
The non-zero fixed points for the endemic and saddle (superscript $sadd$) states satisfy  $p_{i}\!=\!p_{si}\!=\!p_{ii}\!=\!\dot{x}_{i}\!=\!\dot{x}_{si}\!=\!\dot{x}_{ii}\!=\!0$. 

In addition to the forms, Eqs.(\ref{eq:Trans}-\ref{eq:Sadd}), a time-scale must be chosen for a trial solution\cite{Lindley2013}. In general, a good choice is the inverse exponent of the linearized Eqs.(\ref{eq:EOM1}-\ref{eq:EOM6}) associated with the bifurcation, and multiplied by an order-one constant. Once several solutions are computed from trial solutions near bifurcation, one can construct trial solutions for parameters away from bifurcation using local tangent approximations, and bootstrap to arbitrary parameter values. An example MATLAB code is given in the supplementary material of the published version.   
\bibliographystyle{unsrtnat}
\bibliography{sample} 

\end{document}